\documentclass[aps,prd,twocolumn,showpacs,superscriptaddress,nofootinbib,floatfix]{revtex4-2}

\usepackage{amsmath}
\usepackage{amssymb}
\usepackage{graphicx}
\usepackage{bm}
\usepackage{xcolor}
\usepackage[colorlinks=true,citecolor=blue,linkcolor=blue,urlcolor=blue]{hyperref}

\newcommand{\CH}{\mathcal{C}_{H}}
\newcommand{\Dhel}{\mathcal{D}_{H}}
\newcommand{\Csk}{C_{\mathrm{sk}}}
\newcommand{\Cl}{C_{l_{1}}}
\newcommand{\eps}{\epsilon}
\newcommand{\Tr}{\operatorname{Tr}}

\begin{document}

\title{Response of Hellinger-distance based coherence to weak decoherence\\ in two-flavor neutrino oscillations}

\author{Saurabh Rai}
\email{saurabhrai25@iitk.ac.in}
\affiliation{Indian Institute of Technology Kanpur, Department of Physics, \\Uttar Pradesh 208016, India}
\author{Nilakshi Das}
\email{nilakshi.das@iitgn.ac.in}
\affiliation{Indian Institute of Technology Gandhinagar, Department of Physics, \\ Gujarat 382355, India}

\author{Tejhas Kapoor}
\email{kapoor@lpccaen.in2p3.fr}
\affiliation{LPC Caen, Normandie Univ, ENSICAEN, UNICAEN, CNRS/IN2P3, 6 boulevard Maréchal Juin, Caen, 14050 France}

\date{\today}

\begin{abstract}
We evaluate the Hellinger-distance coherence of the two-flavor neutrino state subject to Lindblad damping of the mass-eigenstate interference term, and obtain a closed expression in terms of the flavor-basis density-matrix elements. In the two-flavor vacuum treatment with negligible wave-packet separation, the undamped state is pure, and the smallest eigenvalue of the damped state grows linearly with $\eps=1-\kappa$, where $\kappa=e^{-\Gamma L}$. An expansion about the pure state then yields a nonanalytic $\sqrt{\eps}$ correction to the Hellinger coherence, with coefficient $K=2\sqrt{2}\,u\sin2\theta/\sqrt{1-2u}$, whereas the flavor-transition probability, the $l_1$ coherence, the concurrence, the entanglement of formation and the local quantum Fisher information all respond linearly in $\eps$. Using representative Daya Bay, KamLAND and MINOS working points together with a current $90\%$ C.L. bound on energy-independent damping, we find fractional changes in the Hellinger coherence of $0.18\%$, $9.40\%$ and $21.51\%$, against $0.0009\%$, $0.43\%$ and $1.81\%$ for the concurrence. The comparison concerns the response of the quantifiers to the damping parameter and carries no implication about experimental resolution. The same mechanism operates, with a different coefficient, for a correlated dephasing channel acting on the flavor coherence, where the non-Markovian regime produces damped revivals.
\end{abstract}

\maketitle

\section{Introduction}

Neutrinos produced in weak interactions are associated with a definite flavor, which, under coherent production, corresponds to a superposition of neutrino mass eigenstates. Over a baseline $L$ at energy $E$ the mass eigenstates accumulate different phases, so the probability of detecting a given flavor varies periodically with $L/E$ \cite{Pontecorvo, MNS, Giunti}. An oscillation measurement is in this sense a measurement of the relative phase between mass eigenstates, and is degraded by any process that destroys it. The oscillating neutrino has accordingly become a convenient system in which to evaluate quantities borrowed from quantum information theory, among them flavor entanglement \cite{Blasone2008, Blasone2009}, quantum discord \cite{Wang2023, Dixit2018}, local quantum uncertainty \cite{Girolami2013} and several coherence monotones \cite{Song2018, Ming2020}; the nonclassicality of the flavor evolution has also been probed experimentally through violation of a Leggett--Garg inequality \cite{Formaggio2016}.

Possible non-unitary effects arising from open-system dynamics can be parametrized, under the Markovian approximation, by a Lindblad dissipator. Assuming Lindblad operators diagonal in the neutrino mass basis, the dissipative evolution suppresses the coherence between different mass eigenstates, with the off-diagonal density-matrix elements acquiring a damping factor $\kappa=e^{-\Gamma_{ij}(E)L}$. This is the same coherence term whose suppression reduces the amplitude of the oscillatory component of the survival probability, allowing oscillation data to constrain the corresponding decoherence parameters; the analysis of Ref.~\cite{DeRomeri2023} gives $\Gamma\le5.1\times10^{-24}$~GeV at $90\%$ C.L.\ for energy-independent damping with $\Gamma_{21}=\Gamma_{31}=\Gamma_{32}$, a limit set by MINOS/MINOS+ data. Information-theoretic quantities have been evaluated on states damped in this way, and on states dephased directly in the flavor basis \cite{EPJC2025, Loulijat2026}.

The observation underlying the present work is that the response of such a quantity to weak damping depends qualitatively on how it is constructed from the density matrix. In the two-flavor vacuum treatment, assuming coherent propagation with negligible wave-packet separation, the propagating neutrino state remains pure, and its density matrix therefore has a vanishing eigenvalue. Damping lifts that eigenvalue linearly in $\eps=1-\kappa$. Consequently, quantities that are analytic functions of the density-matrix elements admit expansions in integer powers of $\eps$ with a generically linear leading correction; the survival probability is one such quantity. In contrast, quantities involving $\sqrt{\rho}$ can exhibit nonanalytic behavior at the pure-state boundary. The square root of an eigenvalue that vanishes linearly in $\eps$ contributes at order $\sqrt{\eps}$. Thus, in the weak-damping regime, the decoherence-induced correction to the oscillation probability scales as $\Gamma L$, whereas the corresponding correction to a square-root-based coherence measure scales as $\sqrt{\Gamma L}$. Their ratio of fractional responses therefore scales as $\eps^{-1/2}$ in the weak-damping regime.

We make this quantitative using the Hellinger-distance coherence introduced in Ref.~\cite{JinFei2018}, which satisfies the strong monotonicity requirement of the resource theory of coherence \cite{BCP2014} while retaining a closed expression, and which is related to Yu's skew-information coherence \cite{Yu2017} by a simple functional relation. We derive the measure in closed form for the damped two-flavor state, obtain the $\sqrt{\eps}$ law together with its coefficient, quantify its magnitude at three representative working points using a current bound on $\Gamma$, and show that the same mechanism operates for a correlated dephasing channel acting on the flavor coherence. Throughout, the comparison concerns the response of the quantifiers to the underlying damping parameter. Questions of experimental resolution and of how the density matrix could be reconstructed from data are left for future work.

Section~\ref{sec:framework} sets up the two-qubit description of the state, the damping map, and the coherence measure. Section~\ref{sec:closed} derives the closed form. Section~\ref{sec:sqrt} contains the main result. Section~\ref{sec:bench} evaluates it at three benchmark working points. Section~\ref{sec:memory} treats a second channel as an extension. Section~\ref{sec:concl} concludes. Two appendices collect the explicit matrix elements and the values of the other quantifiers for this state family.

\section{Framework}
\label{sec:framework}

\subsection{The oscillating state as a two-qubit state}

We work in the two-flavor approximation, appropriate when a single mass-squared splitting dominates the channel under consideration. Starting from a flavor eigenstate $|\nu_\alpha\rangle$, the state is a superposition of $|\nu_\alpha\rangle$ and $|\nu_\beta\rangle$ with survival and transition probabilities $P_{\alpha\alpha}$ and $P_{\alpha\beta}$ obeying $P_{\alpha\alpha}+P_{\alpha\beta}=1$. In vacuum and in the absence of damping,
\begin{align}
P_{\alpha\alpha} &= 1-\sin^{2}2\theta\,\sin^{2}\psi , \label{eq:Psurv}\\
\psi &\equiv 1.267\,\frac{\Delta m^{2}[\mathrm{eV}^{2}]\,L[\mathrm{km}]}{E[\mathrm{GeV}]} . \nonumber
\end{align}

Following the occupation-number correspondence of Ref.~\cite{Blasone2008}, $|\nu_\alpha\rangle \equiv |1\rangle_\alpha\otimes|0\rangle_\beta \equiv |10\rangle$ and $|\nu_\beta\rangle \equiv |01\rangle$, the flavor state becomes a two-qubit state whose density matrix, in the ordered basis $\{|00\rangle,|01\rangle,|10\rangle,|11\rangle\}$, is supported entirely on the single-occupancy block,
\begin{equation}
\rho =
\begin{pmatrix}
0 & 0 & 0 & 0\\
0 & p & z^{*} & 0\\
0 & z & q & 0\\
0 & 0 & 0 & 0
\end{pmatrix},
\qquad
\varrho \equiv \begin{pmatrix} p & z^{*}\\ z & q\end{pmatrix},
\label{eq:rho}
\end{equation}
with $p=P_{\alpha\beta}$, $q=P_{\alpha\alpha}$ the populations and $z$ the flavor coherence. 

\subsection{Damping of the mass coherence}

We model neutrino decoherence as dephasing in the mass basis. With
$\rho^{(m)}$ denoting the density matrix in that basis and with the
dephasing Lindblad operator
$L_\phi=\sqrt{\Gamma/2}\,\sigma_z^{(m)}$, the dissipative part of the
master equation is
\begin{equation}
\dot{\rho}^{(m)}
=
\frac{\Gamma}{2}
\left(
\sigma_z^{(m)}\rho^{(m)}\sigma_z^{(m)}
-\rho^{(m)}
\right),
\end{equation}
which gives
\begin{equation}
\rho^{(m)}_{12}(L)
=
\kappa\,\rho^{(m)}_{12}(0)\,
e^{-i\Delta m^{2}L/2E},
\qquad
\kappa=e^{-\Gamma L},
\label{eq:kappa}
\end{equation}
while leaving the mass populations unchanged. Thus, in this
dephasing model, suppression of the mass-basis coherence directly
reduces the oscillatory interference term in $P_{\alpha\alpha}$,
allowing oscillation data to constrain $\Gamma$. Wave-packet
separation provides an additional suppression of the same
interference term; for independent decoherence mechanisms, the
corresponding damping factors combine multiplicatively,
$\kappa_{\rm tot}=\kappa_{\rm Lindblad}\kappa_{\rm wp}$.
Appendix~\ref{app:wp} gives the explicit matrix elements for arbitrary
$\kappa$. We write
\begin{equation}
\eps \equiv 1-\kappa
\end{equation}
for the total damping, which is the small parameter of Sec.~\ref{sec:sqrt}.

Transforming Eq.~\eqref{eq:kappa} to the flavor basis and reading off the elements of Eq.~\eqref{eq:rho} gives, with $c=\cos\theta$, $s=\sin\theta$ and $m\equiv c^{2}s^{2}=\tfrac14\sin^{2}2\theta$,
\begin{align}
p &= 2m\left(1-\kappa\cos2\psi\right), \qquad q=1-p ,
\label{eq:pkappa}\\
|z|^{2} &= m\Big[\cos^{2}2\theta\left(1-2\kappa\cos2\psi\right) \nonumber\\
&\qquad\ \ +\kappa^{2}\left(c^{4}+s^{4}-2m\cos4\psi\right)\Big] .
\label{eq:zkappa}
\end{align}
Both $p$ and $|z|^{2}$ are analytic in $\kappa$, and at $\kappa=1$ Eq.~\eqref{eq:zkappa} reduces identically to $|z|^{2}=pq$, so the undamped state is pure. The determinant of the block takes the simple exact form
\begin{equation}
\det\varrho = pq-|z|^{2} = m\left(1-\kappa^{2}\right),
\label{eq:det}
\end{equation}
which follows immediately from the fact that $\varrho$ is unitarily related to $\rho^{(m)}$, whose determinant is $c^{2}s^{2}(1-\kappa^{2})$.

It is convenient to abbreviate the undamped value
\begin{equation}
u \equiv \left. pq \right|_{\kappa=1} = P_{\alpha\alpha}\left(1-P_{\alpha\alpha}\right) \in \left[0,\tfrac14\right] .
\end{equation}

The chain leading to the result of Sec.~\ref{sec:sqrt} is worth stating at the outset, since its first two links are assumptions:
\begin{align}
&\text{two flavors, vacuum, negligible wave-packet damping} \nonumber\\
&\quad\Longrightarrow\quad \rho\big|_{\kappa=1}\ \text{pure},\qquad \lambda_-\big|_{\kappa=1}=0 \nonumber\\
&\quad\Longrightarrow\quad \lambda_-(\eps) = 2m\,\eps + O(\eps^{2}) \nonumber\\
&\quad\Longrightarrow\quad \CH(\eps) = \CH^{(0)} - K\sqrt{\eps} + O(\eps).
\label{eq:chain}
\end{align}
Section~\ref{sec:validity} discusses what happens when either of the first two conditions is relaxed.

\subsection{Hellinger-distance coherence}

For a fixed reference basis $\{|k\rangle\}$ the incoherent states are $\mathcal{I}=\{\delta=\sum_k \delta_k |k\rangle\langle k|\}$. With the quantum Hellinger distance $D_H(\rho,\delta)=\Tr(\sqrt{\rho}-\sqrt{\delta})^{2}$, it was shown in \cite{JinFei2018} that
\begin{equation}
\CH(\rho) = \min_{\delta\in\mathcal{I}} D_H(\rho,\delta)
= 2\left(1-\sqrt{\textstyle\sum_k \langle k|\sqrt{\rho}|k\rangle^{2}}\,\right)
\label{eq:CHdef}
\end{equation}
is a bona fide coherence measure, satisfying the faithfulness, strong monotonicity and convexity criteria of Ref.~\cite{BCP2014}, with the optimum attained at $\delta^{0}=\sum_k \langle k|\sqrt{\rho}|k\rangle^{2}|k\rangle\langle k|/\sum_{k'} \langle k'|\sqrt{\rho}|k'\rangle^{2}$. The bracket in Eq.~\eqref{eq:CHdef} is related to Yu's skew-information coherence \cite{Yu2017},
\begin{equation}
\Csk(\rho) = \sum_k I\!\left(\rho,|k\rangle\langle k|\right) = 1-\sum_k \langle k|\sqrt{\rho}|k\rangle^{2},
\label{eq:Csk}
\end{equation}
with $I(\rho,K)=-\tfrac12\Tr[\sqrt{\rho},K]^{2}$ the Wigner--Yanase skew information \cite{WY1963,Luo2004}, so that \cite{JinFei2018}
\begin{equation}
\CH = 2\left(1-\sqrt{1-\Csk}\right).
\label{eq:CHCsk}
\end{equation}
Only the square root of $\rho$ is required. That is the feature responsible for the behavior described in Sec.~\ref{sec:sqrt}, and it is also the step that the purity of the undamped state trivializes. Throughout, the reference basis is the flavor (occupation-number) basis.

\section{Closed form for the damped state}
\label{sec:closed}

For a pure state $\sqrt{\rho}=\rho$ and Eq.~\eqref{eq:CHdef} reduces to a function of the populations alone. For non-zero damping and nontrivial mixing, $\rho$ is rank two and that shortcut is lost. The block structure of Eq.~\eqref{eq:rho} nevertheless keeps the problem elementary: the two empty basis states contribute nothing, and only the root of the $2\times2$ block is required.

For any positive $2\times2$ matrix $A$, Cayley--Hamilton gives $A^{2}=(\Tr A)A-(\det A)\,\mathbb{I}$, so that $\left(A+\sqrt{\det A}\,\mathbb{I}\right)^{2} = \left(\Tr A + 2\sqrt{\det A}\right)A$ and
\begin{equation}
\sqrt{A} = \frac{A+\sqrt{\det A}\,\mathbb{I}}{\sqrt{\Tr A + 2\sqrt{\det A}}} .
\label{eq:sqrt2x2}
\end{equation}
Applying this to $\varrho$, with $\Tr\varrho=1$ and
\begin{equation}
\Delta \equiv \sqrt{\det \varrho} = \sqrt{pq-|z|^{2}} = \tfrac12\sin2\theta\,\sqrt{1-\kappa^{2}} ,
\label{eq:Delta}
\end{equation}
the diagonal of $\sqrt{\rho}$ in the full four-dimensional basis is
\begin{equation}
\left(\langle k|\sqrt{\rho}|k\rangle\right)_k
=\left(0,\ \frac{p+\Delta}{\sqrt{1+2\Delta}},\ \frac{q+\Delta}{\sqrt{1+2\Delta}},\ 0\right).
\label{eq:diag}
\end{equation}
The last equality in Eq.~\eqref{eq:Delta} uses Eq.~\eqref{eq:det}; the rest of this section holds for any $p$, $q$, $|z|$.

Because the two vanishing entries contribute $0^{2}$, evaluating $\CH$ on the two-qubit occupation state gives the same number as evaluating it on the bare single-flavor qubit. This property is familiar from the pure case and survives damping unchanged. It does not extend to correlation measures, which are intrinsically bipartite.

Substituting Eq.~\eqref{eq:diag} into Eq.~\eqref{eq:CHdef} and using $p+q=1$ together with $\Delta^{2}=pq-|z|^{2}$, the populations cancel:
\begin{align}
\sum_k \langle k|\sqrt{\rho}|k\rangle^{2}
&= \frac{(p+\Delta)^{2}+(q+\Delta)^{2}}{1+2\Delta} \nonumber\\
&= \frac{1-2pq+2\Delta+2\Delta^{2}}{1+2\Delta}
= 1-\frac{2|z|^{2}}{1+2\Delta} ,
\end{align}
so that
\begin{equation}
\boxed{\ \Csk = \frac{2|z|^{2}}{1+2\sqrt{\det\varrho}}\ ,
\qquad
\CH = 2\left[1-\sqrt{1-\Csk}\ \right] .\ }
\label{eq:CHclosed}
\end{equation}
Equation~\eqref{eq:CHclosed} is exact for every physical state of the form of Eq.~\eqref{eq:rho}, i.e.\ for any $p+q=1$ and $|z|^{2}\le pq$, and therefore applies to any channel that preserves the block structure of Eq.~\eqref{eq:rho}; the mass-basis damping of Sec.~\ref{sec:framework} and the flavor-basis channel of Sec.~\ref{sec:memory} are both covered. The first relation is the closed form of the skew-information coherence for an arbitrary qubit state in its reference basis. In Bloch form it reads $\Csk=(1-\sqrt{1-r^{2}})(r_x^{2}+r_y^{2})/(2r^{2})$, which also follows from the standard qubit expression for the skew information of a Pauli observable together with $\Csk=\tfrac12 I(\varrho,\sigma_z)$~\cite{Yu2017}; the form in terms of $|z|$ and $\det\varrho$ is the one used below.

For mass-basis damping, inserting Eqs.~\eqref{eq:zkappa} and \eqref{eq:det} gives $\CH$ as a function of $\theta$, $\psi$ and $\kappa$ alone. At $\kappa=1$ one has $\Delta=0$ and $|z|^{2}=u$, so
\begin{equation}
\CH^{(0)} \equiv \CH\big|_{\kappa=1} = 2\left(1-\sqrt{1-2u}\right),
\label{eq:CHpure}
\end{equation}
the pure-state result. Since $\Csk\le2|z|^{2}\le2pq\le\tfrac12$, the ceiling
\begin{equation}
\CH \le 2-\sqrt{2}\simeq 0.5858 ,
\end{equation}
attained at equal flavor mixing in the undamped limit, is untouched by the channel. Where a comparison across configurations is intended, we quote the normalized measure $\tilde{\CH}=\CH/(2-\sqrt2)$.

Two features of Eq.~\eqref{eq:CHclosed} matter later. The populations have disappeared from the numerator and survive only inside $\Delta$, so the measure is controlled by the coherence amplitude, with the populations entering through the mixedness they induce. And for small argument $\CH\simeq\Csk$, which is quadratic in $|z|$, whereas the $l_1$ norm $\Cl=2|z|$ is linear. The values of the other quantifiers for this state family are collected in Appendix~\ref{app:others}.

As a numerical consistency check, the closed forms reported here and in Appendix~\ref{app:others} were verified against direct construction of the $4\times4$ density matrix followed by eigendecomposition and explicit matrix-square-root evaluation, over $4\times10^{4}$ random parameter sets, with agreement at the level of $3\times10^{-15}$.

\section{Response to weak damping}
\label{sec:sqrt}

\subsection{Weak-damping expansion}

The eigenvalues of $\varrho$ are $\lambda_\pm = \tfrac12(1\pm R)$ with $R=\sqrt{1-4\det\varrho}$. At $\kappa=1$ the state is pure and $\lambda_-=0$. Using Eq.~\eqref{eq:det} and writing $\kappa=1-\eps$,
\begin{align}
\lambda_- &= m\left(1-\kappa^{2}\right)+O(\eps^{2}) = 2m\,\eps + O(\eps^{2}), \label{eq:lambdaminus}\\
\Delta &= \sqrt{2m\,\eps} + O(\eps^{3/2}). \nonumber
\end{align}
An eigenvalue that lifts off zero linearly makes $\sqrt{\rho}$ nonanalytic at $\eps=0$, and this enters Eq.~\eqref{eq:CHclosed} through the $\Delta$ in the denominator.

The structure of the expansion is general. Suppose a channel gives $\Delta = D\sqrt{\eps}+O(\eps)$ while $|z|^{2}=u+O(\eps)$. Then
\begin{equation}
\Csk(\eps) = 2u - 4uD\sqrt{\eps} + O(\eps),
\end{equation}
and since $\mathrm{d}\CH/\mathrm{d}\Csk = (1-\Csk)^{-1/2}$,
\begin{equation}
\boxed{\ \CH(\eps) = \CH^{(0)} - K\sqrt{\eps} + O(\eps),
\qquad
K = \frac{4uD}{\sqrt{1-2u}} .\ }
\label{eq:sqrtlaw}
\end{equation}
For mass-basis damping $D=\sin2\theta/\sqrt{2}$ by Eq.~\eqref{eq:lambdaminus}, giving
\begin{equation}
K = \frac{2\sqrt{2}\,u\sin2\theta}{\sqrt{1-2u}} .
\label{eq:Kmass}
\end{equation}
Equation~\eqref{eq:sqrtlaw} is the asymptotic expansion of the exact result Eq.~\eqref{eq:CHclosed} about $\eps=0$, not an independent statement; Table~\ref{tab:accuracy} compares the two directly.

The other quantities behave differently. By Eqs.~\eqref{eq:pkappa} and \eqref{eq:zkappa}, $p$ and $|z|^{2}$ are analytic in $\kappa$, so the transition probability and any quantifier that is an analytic function of the matrix elements has an expansion in $\eps$ with a linear leading term. This covers the $l_1$ coherence and the concurrence, $\Cl=\mathcal{C}=2|z|$, the local quantum Fisher information $\mathcal{F}=4|z|^{2}$, the Bell parameter $B_{\max}=2\sqrt{1+\mathcal{C}^{2}}$, and the entanglement of formation, which is smooth in $\mathcal{C}$ at the working points considered, where $\mathcal{C}<1$.\footnote{At maximal mixing $\mathcal{C}=2\sqrt{u}\,(1-\eps)$ exactly, and at $u=1/4$ one has $\mathcal{C}^{(0)}=1$, where $E_F$ is no longer smooth in $\mathcal{C}$; the composition nevertheless gives a linear response, $E_F\simeq1-\eps/\ln2$. The conclusion is therefore unaffected.} The same mechanism as in Eq.~\eqref{eq:sqrtlaw} appears in the other square-root-based quantity considered here: the Hellinger correlation $\Dhel$, which for qubit--qudit states coincides with the local quantum uncertainty \cite{Girolami2013, JinFei2018}, satisfies $\Dhel=2\Csk$ for this state family and so obeys the same law, $\Dhel=4u-8uD\sqrt{\eps}+O(\eps)$, with a coefficient twice that of $\Csk$, i.e.\ $2\sqrt{1-2u}\, K$ in terms of the coefficient of $\CH$.

\begin{table}[t]
\caption{\label{tab:accuracy} The exact closed form, Eq.~\eqref{eq:CHclosed}, compared with its asymptotic expansion, Eqs.~\eqref{eq:sqrtlaw}--\eqref{eq:Kmass}, at the three benchmark working points of Sec.~\ref{sec:bench}. The values of $\eps$ are the upper limits of Table~\ref{tab:bounds}.}
\begin{ruledtabular}
\begin{tabular}{lcccc}
 & $\eps$ & $\CH$ exact & $\CH^{(0)}-K\sqrt{\eps}$ & rel. diff. \\
\hline
Daya Bay & $1.78\times10^{-5}$ & $0.041665$ & $0.041665$ & $1.6\times10^{-5}$\\
KamLAND  & $4.64\times10^{-3}$ & $0.153745$ & $0.153745$ & $3.7\times10^{-6}$\\
MINOS    & $1.88\times10^{-2}$ & $0.427486$ & $0.422390$ & $1.2\times10^{-2}$\\
\end{tabular}
\end{ruledtabular}
\end{table}

\subsection{Relative fractional response}

Comparing fractional changes,
\begin{equation}
\frac{\delta\CH}{\CH^{(0)}} \simeq \frac{K\sqrt{\eps}}{\CH^{(0)}} ,
\qquad
\frac{\delta \Cl}{\Cl^{(0)}} = O(\eps) ,
\end{equation}
so their ratio grows as $\eps^{-1/2}$. At fixed $\eps$ the prefactor $K/\CH^{(0)}$ takes the values $0.414$, $1.380$ and $1.636$ at the Daya Bay, KamLAND and MINOS working points of Sec.~\ref{sec:bench}, so the configurations closer to equal flavor mixing have the larger relative response. Separately, the ratio diverges as $\eps\to0$, and when each configuration is evaluated at its own bound, where $\eps$ differs by three orders of magnitude, this second effect dominates and the shortest baseline shows the largest ratio. Both effects appear in Table~\ref{tab:bounds}.

\begin{figure}[t]
\includegraphics[width=\columnwidth]{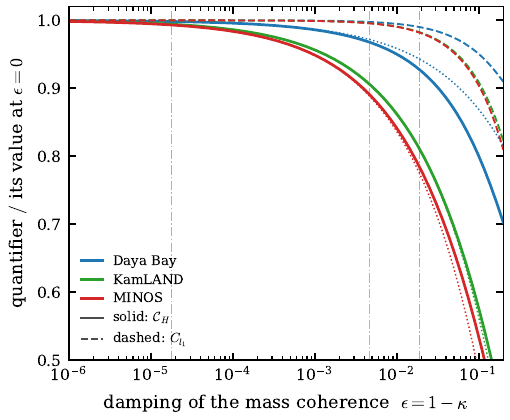}
\caption{\label{fig:sqrtlaw} Fractional response to mass-coherence damping at the three benchmark working points. Solid curves are the exact $\CH(\eps)/\CH^{(0)}$ from Eq.~\eqref{eq:CHclosed}; dotted curves are the asymptotic law $1-K\sqrt{\eps}/\CH^{(0)}$ of Eqs.~\eqref{eq:sqrtlaw}--\eqref{eq:Kmass}; dashed curves are $\Cl(\eps)/\Cl^{(0)}$ for the same configurations, which is linear in $\eps$. Vertical dash-dotted lines mark the upper limits on $\eps$ of Table~\ref{tab:bounds}; the region to the left of each is the weak-damping range allowed by the bound used here, and the region to its right is shown only to expose the crossover.}
\end{figure}

Figure~\ref{fig:sqrtlaw} shows the exact ratio $\CH(\eps)/\CH^{(0)}$ together with the asymptotic law and the corresponding $l_1$ result. The separation between the two behaviors opens rapidly below $\eps\sim10^{-2}$.

\subsection{Scope}
\label{sec:validity}

The square-root behavior requires $\Delta=0$ in the undamped limit, that is, a pure reference state, and both of the first two links of Eq.~\eqref{eq:chain} are needed for this. Since wave-packet separation enters $\kappa$ in the same way as the Lindblad damping, a fixed wave-packet contribution $\kappa_{\rm wp}<1$ means the state is already mixed before $\Gamma$ is switched on; the expansion must then be taken about $\kappa_{\rm wp}$ rather than about unity, $\Delta$ is bounded away from zero, and the response to the additional Lindblad damping is analytic with a slope set by the residual mixedness. The exact expression Eq.~\eqref{eq:CHclosed} covers that case through $\det\varrho$. Throughout we take $\kappa_{\rm wp}=1$. At the reactor working points this is an assumption about the wave-packet width rather than a consequence of data: with the $\kappa_{\rm wp}$ of Appendix~\ref{app:wp}, $\eps_{\rm wp}<\eps_{90}$ requires $\sigma_x\gtrsim4\times10^{-3}$~nm at Daya Bay and $\sigma_x\gtrsim1\times10^{-3}$~nm at KamLAND, whereas at the current lower bound from reactor data, $\sigma_x>2.1\times10^{-4}$~nm~\cite{deGouvea2021}, one has $\eps_{\rm wp}\simeq7\times10^{-3}$ at Daya Bay and the response of $\CH$ to $\Gamma_{90}$ there drops from $0.18\%$ to $0.006\%$. The same check is needed before transporting the law to settings such as astrophysical baselines, where wave-packet separation is not negligible.

Two further restrictions should be stated. Equation~\eqref{eq:sqrtlaw} is an expansion about $\eps=0$ and describes the response to weak damping only; the strongly damped regime is governed by Eq.~\eqref{eq:CHclosed} directly, and there $\CH$ is not in general monotonic in $\kappa$. And the ratio of fractional responses compares two quantifiers evaluated on the same underlying damping parameter within the model adopted here. It is not a statement about the precision with which any of these quantities could be reconstructed from data, which would depend on how the density matrix is accessed and is outside the scope of this work.

\section{Benchmark working points}
\label{sec:bench}

\subsection{Working points}

We adopt three representative configurations spanning the reactor and accelerator sectors, with oscillation parameters and baselines as in Table~\ref{tab:params}. The corresponding energies, $E=4.03$~MeV, $6.07$~MeV and $3.16$~GeV, lie within the quoted experimental envelopes, the last close to the NuMI flux peak.\footnote{The oscillation parameters are those of Ref.~\cite{EPJC2025}, with $\sin^{2}2\theta=0.870$ for KamLAND corresponding to $\tan^{2}\theta_{12}=0.47$, which facilitates comparison with the $l_1$ and entanglement results reported there; that reference quotes baseline and energy ranges, within which our choices of $L$ and $E$ lie.} These are benchmark parameter choices rather than a simulation of the corresponding experiments.

\begin{table*}[t]
\caption{\label{tab:params} Oscillation parameters and benchmark working points. $P_{\alpha\alpha}$ is taken as exact and defines each working point; $L/E$ is the value that reproduces it through Eq.~\eqref{eq:Psurv}. $u=P_{\alpha\alpha}(1-P_{\alpha\alpha})$ is the undamped value and $\CH^{(0)}$ follows from Eq.~\eqref{eq:CHpure}. The baseline $L$ is the one used in Table~\ref{tab:bounds}.}
\begin{ruledtabular}
\begin{tabular}{lccccccc}
 & $\Delta m^{2}$ [eV$^{2}$] & $\sin^{2}2\theta$ & $L/E$ & $L$ [km] & $P_{\alpha\alpha}$ & $u$ & $\CH^{(0)}$ \\
\hline
Daya Bay & $2.42\times10^{-3}$ & $0.084$ & $0.171$\footnotemark[1] & $0.69$ & $0.9789$ & $0.02065$ & $0.0417$\\
KamLAND  & $7.49\times10^{-5}$ & $0.870$ & $29.7$\footnotemark[1]  & $180$  & $0.9108$ & $0.08125$ & $0.1697$\\
MINOS    & $2.32\times10^{-3}$ & $0.950$ & $232$\footnotemark[2]   & $735$  & $0.6215$ & $0.23524$ & $0.5446$\\
\end{tabular}
\end{ruledtabular}
\footnotetext[1]{in km/MeV}
\footnotetext[2]{in km/GeV}
\end{table*}

The last column of Table~\ref{tab:params} gives the undamped coherence, the MINOS entry reaching $93\%$ of the ceiling $2-\sqrt{2}$. Normalized to Daya Bay, the ratios across configurations are $1:1.98:3.38$ for $\Cl$, $1:2.94:6.49$ for $E_F$ and $1:4.07:13.05$ for $\CH$, the last being more pronounced because $\CH$ is quadratic in the coherence amplitude where $\Cl$ is linear. The differences among the benchmark points should not be read as a direct baseline dependence of coherence; at the chosen working points they reflect the different oscillation amplitudes encoded in $u$, which is controlled by the mixing angle and by the position along the oscillation.

\begin{figure*}[t]
\includegraphics[width=\textwidth]{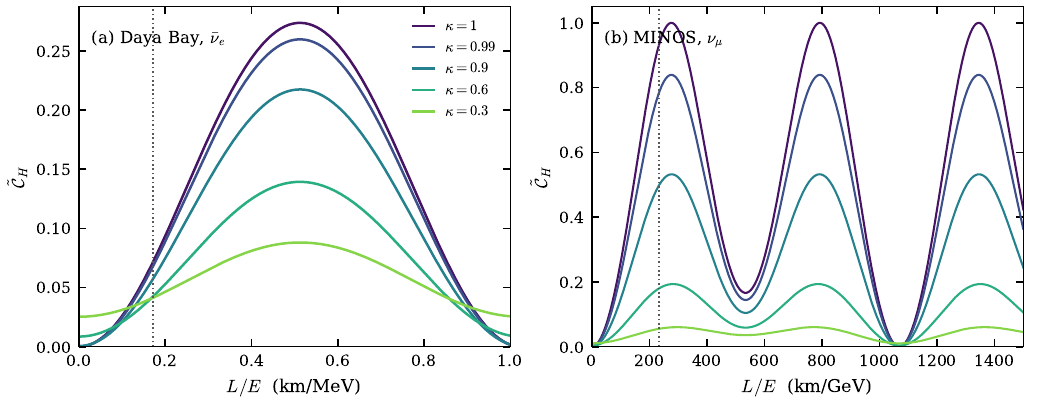}
\caption{\label{fig:LE} Normalized Hellinger coherence $\tilde{\CH}=\CH/(2-\sqrt{2})$ against $L/E$ for (a) Daya Bay reactor $\bar\nu_e$ disappearance and (b) MINOS accelerator $\nu_\mu$ disappearance, at several values of the damping factor $\kappa$. Parameters as in Table~\ref{tab:params}; vertical dotted lines mark the benchmark working points. Note the difference in vertical scale. Here $\kappa$ is held fixed along each curve to display the functional dependence; physically $\kappa=e^{-\Gamma L}$ tends to unity as $L\to0$, so the small-$L/E$ region of the curves with $\kappa<1$ is illustrative only.}
\end{figure*}

Figure~\ref{fig:LE} shows the $L/E$ dependence at several values of $\kappa$. The Daya Bay oscillation is shallow, the survival probability dipping only to $0.916$, so the coherence remains well below its ceiling across the accessible range. At MINOS the mixing is nearly maximal; the coherence executes complete cycles, reaching $2-\sqrt{2}$ twice per period, where $P_{\alpha\alpha}=1/2$ (near, but not exactly at, the quarter-period points, since $\sin^{2}2\theta<1$), and returning to zero at the survival nodes where $P_{\alpha\alpha}=1$. Damping reduces the amplitude of these cycles and fills in the nodes, since for $\kappa<1$ the flavor coherence no longer vanishes when the survival probability returns to unity. In the limit $\kappa\to0$ the mass density matrix becomes $\mathrm{diag}(c^{2},s^{2})$, which is not diagonal in the flavor basis unless $\theta=0$ or $\pi/4$, so a residual flavor coherence survives complete damping; at the three working points $\CH(\kappa=0)=0.030$, $0.030$ and $0.012$.

\subsection{Response at currently allowed damping}

To estimate the size of the effect at phenomenologically relevant parameter values, we use the $90\%$ C.L. upper bound on the energy-independent damping parameter obtained in Ref.~\cite{DeRomeri2023} for equal damping of all three mass-basis coherences ($\Gamma_{21}=\Gamma_{31}=\Gamma_{32}\equiv\Gamma$, their Model~A), $\Gamma_{90}=5.1\times10^{-24}$~GeV, set by MINOS/MINOS+ data and corresponding to $2.58\times10^{-5}$~km$^{-1}$, which through $\kappa=e^{-\Gamma L}$ gives the maximum damping listed in Table~\ref{tab:bounds}. This bound assumes an energy-independent $\Gamma$ that is the same for all mass-state pairs, which is what allows a single value to be used at the $\Delta m^{2}_{31}$- and $\Delta m^{2}_{21}$-driven working points alike. Other flavor structures give weaker limits (e.g.\ $3.2\times10^{-23}$~GeV when only $\Gamma_{21}$ is nonzero), other assumed dependences $\Gamma_{ij}(E)=\Gamma_{ij}(E_0)(E/E_0)^{n}$ give different limits, and $\Gamma$ is in any case channel- and experiment-dependent. Hence, the entries of Table~\ref{tab:bounds} are benchmarks rather than universal statements. What matters here is only that the allowed $\eps$ is small enough for the weak-damping expansion of Sec.~\ref{sec:sqrt} to be the relevant regime, and that conclusion does not depend on the choice.


\begin{table*}[t]
\caption{\label{tab:bounds} Fractional change $|\delta X/X|$ of each quantifier at the $90\%$ C.L. upper bound on the energy-independent damping parameter of Ref.~\cite{DeRomeri2023}, evaluated at the baseline of Table~\ref{tab:params}. All entries are decreases except those of $P_{\alpha\beta}$, which increases at each of these working points.}
\begin{ruledtabular}
\begin{tabular}{lcccccccc}
 & & \multicolumn{2}{c}{square-root response, $O(\sqrt{\eps})$} & \multicolumn{5}{c}{analytic response, $O(\eps)$}\\
\cline{3-4}\cline{5-9}
 & $\eps_{90}$ & $\CH$ & $\Dhel$ & $\Cl=\mathcal{C}$ & $E_F$ & $\mathcal{F}=\mathcal{C}^{2}$ & $B_{\max}$ & $P_{\alpha\beta}$\\
\hline
Daya Bay & $1.78\times10^{-5}$ & $0.18\%$  & $0.17\%$  & $0.0009\%$ & $0.0015\%$ & $0.0019\%$ & $0.0001\%$ & $0.0018\%$\\
KamLAND  & $4.64\times10^{-3}$ & $9.40\%$  & $9.03\%$  & $0.43\%$   & $0.66\%$   & $0.86\%$   & $0.11\%$   & $1.80\%$\\
MINOS    & $1.88\times10^{-2}$ & $21.51\%$ & $18.85\%$ & $1.81\%$   & $2.60\%$   & $3.58\%$   & $0.87\%$   & $0.48\%$\\
\end{tabular}
\end{ruledtabular}
\end{table*}

\begin{figure}[t]
\includegraphics[width=\columnwidth]{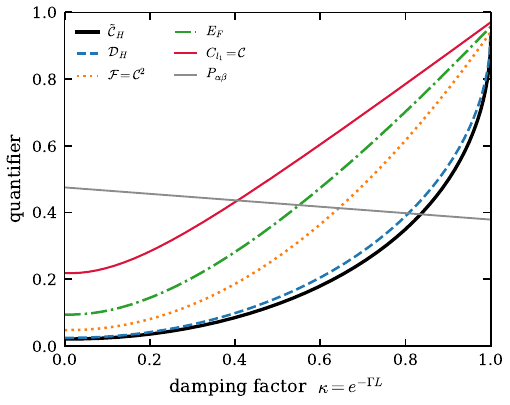}
\caption{\label{fig:quantifiers} Quantifiers at the MINOS working point as functions of the damping factor $\kappa=e^{-\Gamma L}$: the normalized Hellinger coherence $\tilde{\CH}$, the Hellinger correlation $\Dhel$, the local quantum Fisher information $\mathcal{F}=\mathcal{C}^{2}$, the entanglement of formation $E_F$, the $l_1$ coherence, which for this state equals the concurrence, and the transition probability $P_{\alpha\beta}$. The analytic and square-root responses separate near $\kappa=1$, where $\tilde{\CH}$ and $\Dhel$ approach their undamped values with infinite slope; away from that neighborhood, the figure shows only the global shape of each quantifier and implies no general classification.}
\end{figure}

Over the whole allowed range the concurrence and the transition probability move by under two percent, and the other analytic quantifiers by at most a few percent ($3.6\%$ for $\mathcal{F}=\mathcal{C}^{2}$ and $2.6\%$ for $E_F$, both at MINOS), while the Hellinger coherence responds by up to a fifth of its value. The contrast is between an $O(\eps)$ and an $O(\sqrt{\eps})$ response to the same underlying parameter. Figure~\ref{fig:quantifiers} shows the full $\kappa$ dependence of the quantifiers at the MINOS working point, together with $P_{\alpha\beta}$ itself.

\section{Extension: a flavor-basis channel with memory}
\label{sec:memory}

Equation~\eqref{eq:CHclosed} holds for any channel preserving the block structure of Eq.~\eqref{eq:rho}, so it is straightforward to ask whether the mechanism of Sec.~\ref{sec:sqrt} is particular to mass-basis damping. As an extension we take the correlated dephasing channel of Ref.~\cite{EPJC2025}, which acts directly on the flavor coherence,
\begin{equation}
z \longrightarrow \eta(t)\,z ,
\qquad
\eta(t) = f^{2}(t) + \left[1-f^{2}(t)\right]\gamma_c ,
\label{eq:etat}
\end{equation}
leaving $p$ and $q$ unchanged, with $\gamma_c\in[0,1]$ the strength of the classical correlation between successive interactions and $f(t)$ the decoherence function of random-telegraph noise,
\begin{equation}
f(t)=e^{-t/2\tau}\left[\cosh\frac{vt}{2\tau}+\frac{1}{v}\sinh\frac{vt}{2\tau}\right],
\quad v=\sqrt{1-16\tau^{2}} ,
\label{eq:fRTN}
\end{equation}
for $\tau<1/4$, and its analytic continuation
\begin{equation}
f(t)=e^{-t/2\tau}\left[\cos\frac{v't}{2\tau}+\frac{1}{v'}\sin\frac{v't}{2\tau}\right],
\quad v'=\sqrt{16\tau^{2}-1} ,
\end{equation}
for $\tau>1/4$. The first regime is memoryless and $f$ decays monotonically; the second retains memory, $f$ oscillates, and information flows back from the environment. This is a different physical channel from that of Sec.~\ref{sec:framework}: here $P_{\alpha\beta}$ is unchanged by construction, and the bound of Table~\ref{tab:bounds} does not apply to $\eta$.

For this channel $\Delta=\sqrt{u\left(1-\eta^{2}\right)}$, so with $\eps=1-\eta$ one has $D=\sqrt{2u}$ in Eq.~\eqref{eq:sqrtlaw} and
\begin{equation}
K_{\rm flavor} = \frac{4\sqrt{2}\,u^{3/2}}{\sqrt{1-2u}} ,
\end{equation}
which equals $0.017$, $0.143$ and $0.887$ at the three working points, against $0.017$, $0.234$ and $0.891$ for mass-basis damping. The $\sqrt{\eps}$ law therefore survives the change of channel, with a coefficient that depends on it. This is as expected from Eq.~\eqref{eq:chain}: what is required is only that the undamped state be pure and that the smallest eigenvalue grow linearly.

\begin{figure*}[t]
\includegraphics[width=\textwidth]{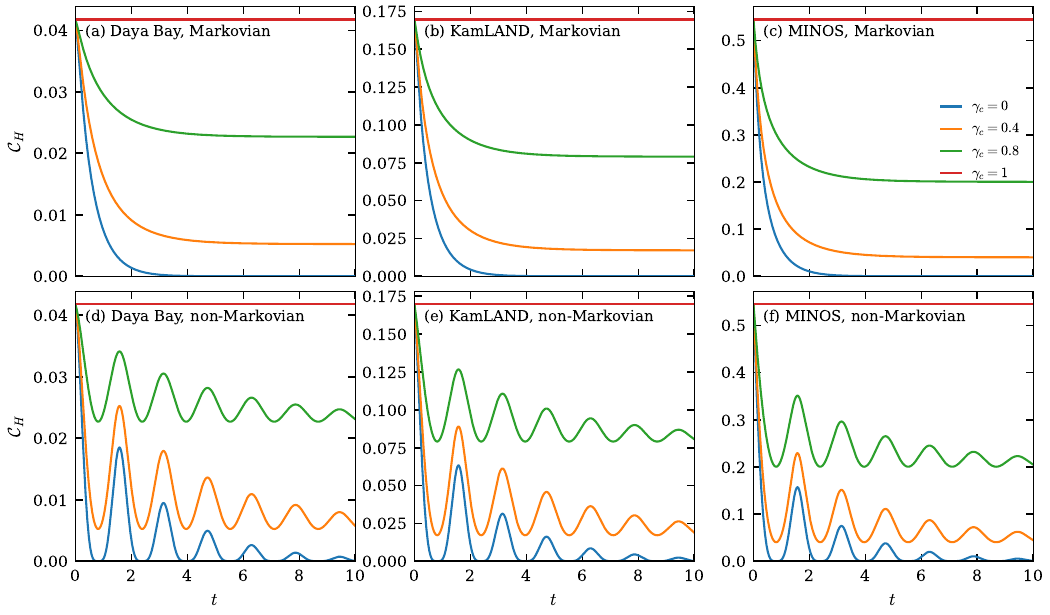}
\caption{\label{fig:dynamics} Hellinger coherence under the flavor-basis correlated dephasing channel of Eq.~\eqref{eq:etat}, in the memoryless regime $\tau=0.1$ (top row) and the memory-retaining regime $\tau=5$ (bottom row), for classical correlation strengths $\gamma_c=0,\,0.4,\,0.8,\,1$. Columns correspond to the three working points of Table~\ref{tab:params}. The values $\gamma_c=0.4$ and $0.8$ are chosen to illustrate the channel dynamics and correspond to asymptotic coherence losses far outside the weak-damping region of Table~\ref{tab:bounds}. Note the different vertical scales.}
\end{figure*}

Figure~\ref{fig:dynamics} shows $\CH(t)$ for the three working points in both noise regimes. In the memoryless case, the decay is monotonic to the plateau obtained from Eq.~\eqref{eq:CHclosed} at $\eta=\gamma_c$, which grows roughly as $\gamma_c^{2}$: raising $\gamma_c$ from $0.4$ to $0.8$ doubles $\Cl$ but multiplies $\CH$ by $4.3$, $4.6$ and $5.0$. At $\gamma_c=1$ the channel is completely correlated, $\eta\equiv1$, and the coherence is frozen. In the memory-retaining case the decay acquires damped oscillations; since $\eta$ depends on $f^{2}$, the revivals recur with period $2\pi\tau/v'$ ($\simeq1.57$ for $\tau=5$), half the period of $f$ itself. For $\gamma_c=0$, the function $f$ has isolated zeros at which $\CH$ vanishes exactly before reviving; for $\gamma_c>0$ the minima are lifted and no zero occurs. In neither case does the coherence vanish over a finite interval, in contrast to the nonlocal advantage of quantum coherence, which is known to vanish over finite windows of $L/E$ in this system \cite{Ming2020}. The revivals are weaker in $\CH$ than in the analytic measures: at the first revival maximum with $\gamma_c=0$, $\Cl$ recovers to $73\%$ of its initial value at every working point, while $\CH$ recovers to $44\%$, $37\%$ and $29\%$, again reflecting its quadratic dependence on the coherence amplitude.

This section illustrates that the closed form of Sec.~\ref{sec:closed} evaluates directly for a time-dependent coherence and that the square-root mechanism is not tied to one channel. The weak-damping result of Sec.~\ref{sec:sqrt} does not rely on this particular memory model.

\section{Conclusions}
\label{sec:concl}

We have computed the Hellinger-distance coherence of the two-flavor neutrino state after damping, obtaining the exact closed form Eq.~\eqref{eq:CHclosed}, which depends only on the flavor-basis populations and coherence and therefore applies to any channel preserving the single-occupancy block. The derivation uses that block structure, which reduces the matrix square root to a $2\times2$ problem.

In the two-flavor vacuum treatment with negligible wave-packet separation, the undamped state is pure, its density matrix has a vanishing eigenvalue, and Lindblad damping of the mass coherence lifts that eigenvalue linearly in $\eps=1-e^{-\Gamma L}$. The square root then makes the coherence nonanalytic, giving $\CH=\CH^{(0)}-K\sqrt{\eps}+O(\eps)$ with $K=2\sqrt{2}u\sin2\theta/\sqrt{1-2u}$. The transition probability itself, the $l_1$ norm, the concurrence, the entanglement of formation, the Bell parameter and the local quantum Fisher information all retain analytic expansions with linear leading terms, so the ratio of fractional responses grows as $\eps^{-1/2}$. At the $90\%$ C.L. bound on energy-independent damping the Hellinger coherence at the MINOS working point changes by $21.5\%$ against $1.8\%$ for the concurrence and $0.5\%$ for the transition probability. The Hellinger correlation, equivalently the local quantum uncertainty, follows the same law with coefficient $2\sqrt{1-2u}\, K$, and a flavor-basis correlated dephasing channel reproduces it with a different coefficient, so the behavior is a property of the square root rather than of a particular channel.

The Hellinger coherence is convenient in this context because it is a coherence monotone defined with respect to the flavor basis and has a closed form for the state family considered. The result relies on the pure-state limit of the two-flavor vacuum treatment. Extensions to three-flavor propagation, matter effects and appreciable wave-packet separation would all modify the rank structure of the state and would therefore provide useful tests of the robustness of the square-root response.

\begin{acknowledgments}
S.R. acknowledges fellowship from IIT Kanpur. N.D. acknowledges financial support from the SERB grant SPG/2022/001238.
\end{acknowledgments}

\appendix

\section{Flavor-basis matrix elements}
\label{app:wp}

The elements of Eq.~\eqref{eq:rho} follow from
\begin{align}
F^{\alpha}_{\beta\gamma}&=\sum_{j,k} U^{*}_{\alpha j}U_{\alpha k}\,f_{jk}\,U_{\beta j}U^{*}_{\gamma k}, \nonumber\\
f_{jk}&=\kappa\,e^{\mp2i\psi}\ \ (j\neq k),
\end{align}
with $f_{jj}=1$, $p=F^{\alpha}_{\beta\beta}$, $q=F^{\alpha}_{\alpha\alpha}$ and $z=F^{\alpha}_{\alpha\beta}$. Carrying out the sums gives Eqs.~\eqref{eq:pkappa} and \eqref{eq:zkappa}. Two contributions enter $\kappa$ multiplicatively. Lindblad damping of the mass coherence gives $\kappa_{\rm L}=e^{-\Gamma_{ij}(E)L}$, Eq.~\eqref{eq:kappa}. Wave-packet separation gives
\begin{equation}
\kappa_{\rm wp}=\exp\left[-\left(\frac{\Delta m^{2}x}{4\sqrt{2}E^{2}\sigma_x}\right)^{2}\right],
\end{equation}
so that $\kappa=\kappa_{\rm L}\kappa_{\rm wp}$. Equations~\eqref{eq:pkappa} and \eqref{eq:zkappa} satisfy $p+q=1$ for any $\kappa$, and at $\kappa=1$ reduce to $|z|^{2}=pq$, the pure case. The determinant is $\det\varrho=c^{2}s^{2}(1-\kappa^{2})$ for all $\kappa$, Eq.~\eqref{eq:det}.

\section{Other quantifiers for this state family}
\label{app:others}

For the state of Eq.~\eqref{eq:rho}, direct evaluation gives
\begin{align}
\Cl &= 2|z|, \\
\mathcal{C} &= 2|z| = \Cl, \\
E_F &= h\!\left(\tfrac12\left[1+\sqrt{1-\mathcal{C}^{2}}\right]\right), \\
\mathcal{F} &= 4|z|^{2} = \mathcal{C}^{2}, \\
\Dhel &= \frac{4|z|^{2}}{1+2\Delta} = 2\,\Csk, \\
B_{\max} &= 2\sqrt{1+\mathcal{C}^{2}},
\end{align}
with $\mathcal{C}$ the concurrence and $h(x)=-x\log_2 x-(1-x)\log_2(1-x)$ the binary entropy \cite{Wootters1998}. Here $\mathcal{F}$ is the local quantum Fisher information in the form $1-\lambda_{\max}(M)$, with $M_{\nu\mu}=\sum_{i,j}\frac{2q_iq_j}{q_i+q_j}\langle\psi_i|\sigma_\nu\otimes I|\psi_j\rangle\langle\psi_j|\sigma_\mu\otimes I|\psi_i\rangle$ \cite{Girolami2014}, and $\Dhel$ is the Hellinger correlation of Ref.~\cite{JinFei2018}, equal for qubit--qudit states to the local quantum uncertainty \cite{Girolami2013}; the expression given here agrees with the result of Ref.~\cite{Loulijat2026}. The concurrence follows from the eigenvalues of $\rho\tilde{\rho}$, which are $\vartheta_\pm=(\sqrt{pq}\pm|z|)^{2}$ and $\vartheta_3=\vartheta_4=0$; since $|z|\le\sqrt{pq}$ one obtains $\mathcal{C}=2|z|$. The Bell parameter follows from the Horodecki criterion \cite{Horodecki1995}: because the state occupies only the single-occupancy block, $T_{zz}=-1$ and $T^{\dagger}T=\mathrm{diag}(4|z|^{2},4|z|^{2},1)$, whence $\mathcal{N}=\max(2\mathcal{C}^{2},1+\mathcal{C}^{2})=1+\mathcal{C}^{2}$.\footnote{Taken at face value this implies a Clauser--Horne--Shimony--Holt violation for any $\mathcal{C}>0$. The $T_{zz}=-1$ responsible for it reflects the single-particle sector, in which exactly one flavor mode is occupied, so the measurements involved superpose different occupation numbers of a single mode. Whether that is operationally meaningful for neutrinos raises questions of mode entanglement and superselection that we do not address; $B_{\max}$ is used here only as an example of a quantifier with an analytic response.} For completeness, we note that some of these expressions differ from those printed in Ref.~\cite{EPJC2025} for the same state; the values listed here follow from direct evaluation and are the ones used in Sec.~\ref{sec:bench}.

The relation $\Dhel=2\Csk$ is not accidental. For a qubit, Yu \cite{Yu2017} showed that the skew information of a nondegenerate observable with eigenvalues $\tfrac12\Tr K\pm\lambda$ equals $2\lambda^{2}\Csk$, and the local quantum uncertainty minimizes exactly that over local observables of unit spectrum; the relation holds whenever the minimizing local basis coincides with the reference basis, as it does here. Combining with Eq.~\eqref{eq:CHCsk},
\begin{equation}
\CH = 2\left(1-\sqrt{1-\tfrac12\Dhel}\right),
\label{eq:CHDH}
\end{equation}
so the Hellinger distance supplies a coherence and a correlation diagnostic that remain in one-to-one correspondence for this state family. They are not numerically equal: Eq.~\eqref{eq:CHDH} gives $\CH<\Dhel$ for all $\Dhel>0$, and in the undamped limit $\Dhel=4u$ while $\CH=2(1-\sqrt{1-2u})$, which coincide only at $u=0$.

\end{document}